# Generalized entropies and open random and scale-free networks

Vladimir Gudkov and Vladimir Montealegre

*Department of Physics and Astronomy, University of South Carolina, Columbia, SC 29208*

**Abstract.** We propose the concept of open network as an arbitrary selection of nodes of a large unknown network. Using the hypothesis that information of the whole network structure can be extrapolated from an arbitrary set of its nodes, we use Rényi mutual entropies in different $q$-orders to establish the minimum critical size of a random set of nodes that represents reliably the information of the main network structure. We also identify the clusters of nodes responsible for the structure of their containing network.

One important topic of interest in the study of complex networks is to find a correlation between the structure of the whole network and a representative part of it (a set of randomly selected or chosen nodes). For instance, in the framework of social networks, choosing a sample of people and their links could provide important information about the structure of the large unknown network to which they belong, and for which a complete description might not be feasible due to the commonly large size of complex networks or another possible reason. Therefore, the development of a method that can characterize the whole network from the incomplete information available about it is a helpful tool for the analysis of network vulnerability, topology and evolution [1]. It has been observed that many real world networks like the Internet, Metabolic Networks, the Hollywood actors network and Research Collaborations are scale-free networks [2] that follow a power law degree distribution and for which a high degree of self similarity is expected. Therefore, in this paper we will focus our attention on this particular kind of structure; however the results are applicable to almost any network topology.

The type of network analysis which considers not only node connectivity, but also the network's structure in the attempt to represent the whole network with only a part of it, requires an approach similar to the one used in statistical physics where the entropy function, being defined on a subsystem, contains information about the macroscopic state of the whole system. We describe networks by the adjacency (connectivity) matrix $C$. Then using the assignment of a probability for each node [3, 1] as $p_k = \sum_{i=1}^{n} C_{ik} / \sum_{i,j=1}^{n} C_{ij}$, one can define the mutual entropy of the network as

$$H_q(C) = H_q(P(row)) + H_q(P(column)) - H_q(P(column)|P(row)), \qquad (1)$$

where

$$H_q(P) = \frac{1}{1-q} \log \sum_{j=1}^{n} (p_i)^q \qquad (2)$$

and

$$H_q(P(column)|P(row)) = \frac{1}{1-q} \log(\sum_{i,j}^{n} C_{ij}^q) \qquad (3)$$

are the usual one dimensional and conditional Rényi entropies (see for detail [4,3,1]). Here $n$ is the number of nodes in the network, P is a vector of probabilities $p_i$, and $q>0$ is an order of Rényi entropy (Rényi's entropy is equal to Shannon's entropy in the limit when $q=1$).

We use the concept of mutual entropy to calculate the amount of information contained in a set of interconnected nodes, which can be either the whole network or a selected sample of it.

Being a (not countable) set of functions that includes Shannon's definition of entropy, Rényi entropy supplies a more complete description of the network structure with the important feature that it depicts a more refined measure of the (sub) network's level of disorder as the value of the $q$-order becomes larger. It has also been observed that Rényi entropy can be considered as a measure of localization in complex systems, therefore, if there is redundant information associated to the network structure, we can suggest two hypotheses: the set of entropies can be used to characterize the main properties related to the structure (topological and information exchange) and dynamics of networks; and that the mutual entropies, calculated over a part of the network contain enough information to represent the whole network. These two conjectures are the core of our proposal of the network's structural analysis when the available data is not complete; to perform such type of analysis, certain requirements on the acquirable information must be established. To this end, let us define the concept of an open network as an arbitrary chosen subset of nodes that belong to a large unknown network, thus, in the study of real world networks, the set of available data obtained can be considered to be an open network whose entropy measurements can be "extrapolated" to be a reasonably accurate measure of the whole network.

A crucial requirement to demand from the open network is the minimum critical size it must have to represent the whole network without a significant loss of information related to the main structure. Knowledge of this minimal size permits the definition of a representative sub-network as an open network whose size is larger than the critical one. To find this threshold, let us consider a simulated scale-free network with size 5000 nodes (the size is chosen to avoid possible systematic errors in simulations, see [1]), and from it, let us take randomly chosen sub sets of different sizes (random open networks), thus, by selecting a reasonably large number of these open networks at each size (s), the average mutual entropy $H_q$ can be found with its corresponding uncertainty $\sigma_q(s)$ (Fig. 1a). Different $q$-orders in the figure are represented by q=0 (triangles), $q=1$ (circles), $q=2$ (squares) and the entropy difference between $q=1$ and $q=2$ (stars); each data point in the plot is the average over 100 open networks.

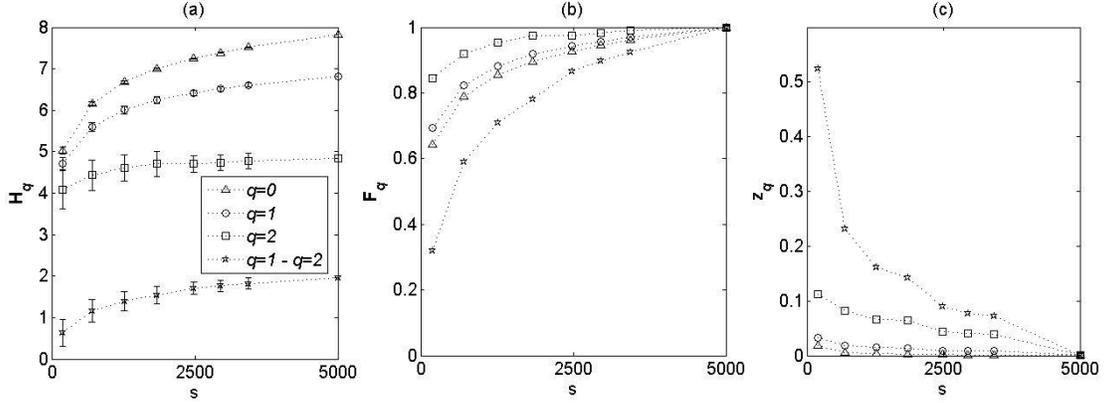

**FIGURE 1.** The plots of (a) The Mutual entropy $H_q$, (b) The rescaled mutual entropy $F_q$ and (c) The relative entropy uncertainty $z_q$ versus the size of the open network. The triangles, circles and squares represent the q-degree values 0, 1, 2 respectively, and the stars represent the difference between q-degrees 1 and 2.

It can be observed that for $q=0,1,2$ the mutual entropy $H_q$ increases rapidly with the size, and after half of the whole network's size has been reached, the entropy value settles in a range near the value of the entropy of the whole network, this can be seen better in fig. 1b where the entropies have been rescaled with respect to the entropy of the whole network by defining $F_q(s)=H_q(s)/H_q(5000)$. For $s>2500$ the rescaled entropies are $F_q>0.9$. It should be noted that when the value of $q$ becomes larger, it gets increasingly difficult to determine the critical size by just observing the rescaled entropies (since the plots become flatter), however, as the value of q increases, the uncertainties in the entropies $\sigma_q(s)$ also become more noticeable (this is consistent with the fact that higher orders of the mutual entropy enhance the contribution of nodes with large degree) in a way such that for each $q$-order, the uncertainties become smaller as the size $s$ increases, thus, $\sigma_q$ is another parameter useful to decide the critical size. We define the relative uncertainty at each open network size as $z_q(s)=\sigma_q(s)/H_q(s)$ (see fig. 1c), the calculations show that $z_{q=0,1,2}<0.05$ for $s>2500$, which is acceptable and agrees with the threshold obtained by observing $F_q$, therefore, the critical size of the open networks is about a half of the size of the whole network.

The nodes of each subset used to create the plots in fig. 1 are chosen randomly, so the mutual entropies $H_q$ found for each size establish a maximal entropy value that should be considered as a worst case in the choice of the size of a representative network. However, if the nodes of the sub sets are chosen according to a selection criterion, the structural information contained in the sample may represent the whole network with a much better fidelity, and this should be observable in the mutual entropy plot of the selected sub sets. To observe this, we rearrange the nodes according to the network cluster structure based on their connectivity. Since the clustering process used should be capable of resolving hierarchical structures, we have used a physics based clustering algorithm [3] which represents the nodes in an initial state with perfectly symmetric conditions (all the *n* nodes are equidistant over a *n-1* dimensional sphere) and allows them to interact via attractive (repulsive) forces according to whether each pair of nodes is connected (disconnected); after the nodes have condensed in the *n-1* dimensional space, the sequence of the nodes in connectivity matrix can be reorganized to show the clusters, as in fig. 2 where the result of this process has been performed over the same matrix used to create fig. 1. We identified six clusters, and it can be seen that they are arranged in a way such that the denser clusters contain the sparser ones, with inter-cluster connections which indicate that the nodes are ordered hierarchically even inside each cluster. The density of cluster 6 is noticeably smaller than the one seen in clusters 1-5, this might lead to the naïve interpretation that cluster 6 is structurally different of clusters 1-6, however, the structure of cluster 6 is still scale-free, as can be seen in fig 3, where the degree distribution of cluster 6 indeed follows a power law, but differs with the whole network in that it does not contain nodes with such a high degree of connectivity, which are found in clusters 1-5. Therefore, large structural components in that cluster are not expected, moreover, it is seen that the cluster 6 covers a very large fraction of the nodes of whole network and it's size is much larger than the sizes of the other identified clusters, this indicates that an open network of any size (randomly chosen) is very likely to contain a significant number of nodes in cluster 6 and this reduces the probability of "fishing" strongly structural nodes in a random sample, which justifies the argument of fig. 1 representing the worst case in the choice of nodes.

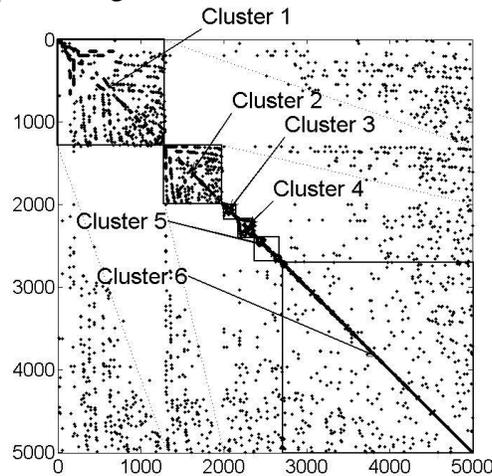

**FIGURE 2.** The self-similar structure of a scalefree network with 5000 nodes is observable in the adjacency matrix. Six clusters (highly structural sub networks) are identified.

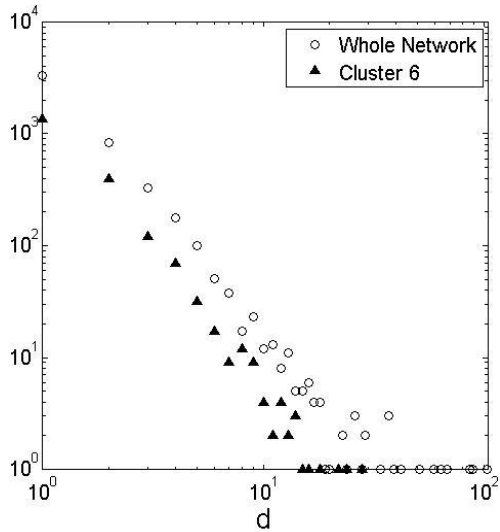

**FIGURE 3.** Degree distribution of the whole 5000 nodes network compared to that of cluster 6

To compare the contributions of each cluster to the network's organization, we plot their mutual entropies for order $q=2$ (see fig. 4a), it can be seen that clusters 1, 2, (1+2), (1+2+3+4+5) are aligned with the mutual entropy of the whole network, while every other combination of clusters possesses a mutual entropy that increases rapidly with the size of the cluster considered (same behavior as a random open network), in particular, clusters 6 (plus sign) and the joint cluster 3+4+5+6 (downwards hollow triangle) have mutual entropies which are even larger than the one of the whole network and the averages of fig. 1a. This can be attributed to the fact that clusters composed of nodes with low degree of connectivity are expected to possess small substructures and therefore they are expected to have a much larger amount of disorder than the one contained in a hierarchical structure.

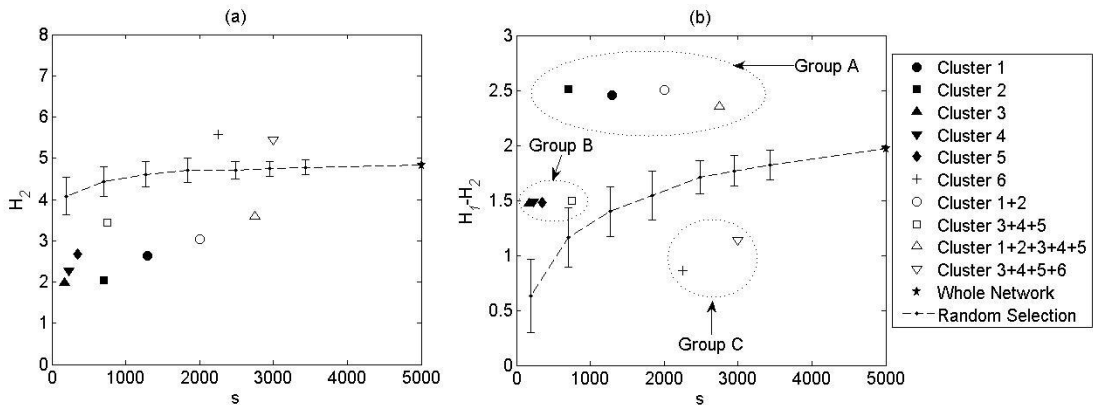

FIGURE 4. Plot of the mutual entropies of (a) order two $H_2$ and (b) the difference of the mutual entropies $H_1$ and $H_2$ for the clusters identified in a scalefree network with 5000 nodes.

Figure 3b shows a plot of the difference of the mutual entropies $\Delta H(s) = H_1(s) - H_2(s)$, three groups of values for $\Delta H$ are found, group A contains clusters 1, 2, (1+2) and (1+2+3+4+5) with $\Delta H \approx 2.5$, group B contains clusters 3, 4, 5 and (3+4+5) with $\Delta H \approx 1.5$ and group C contains clusters 6 and (3+4+5+6) with $\Delta H \leq 1.5$. The mutual entropy difference found for the whole network is $\Delta H \approx 2.0$, which is also the highest value observed for the average of $\Delta H$ calculated for random open networks. This indicates that clusters containing high connectivity degree nodes are more prone to possess a well defined interconnected structure that is enhanced for larger $q$-orders, and therefore they are associated to a larger value of $\Delta H$ which can indeed exceed the entropy difference of the whole network On the other hand, clusters containing nodes with lower connectivity degrees must be made up of small substructures, this manifests in a significantly larger value of the mutual entropy of order $q$=2 (group C) and results in a value of $\Delta H$ which can be smaller than the average for randomly selected open networks. This indicates that a structural subset of nodes in the network can be recognized if it has a significantly larger value of $\Delta H$ than a randomly chosen set of nodes.

In conclusion, we have proposed a method to determine the value of the minimum critical size that an open network must have in order to represent the whole network in a reliable way; we have found this threshold to be about of half of the size of the network for the scale-free case. We also showed that the main topological features of the scale-free network type of structure can be found in certain clusters of nodes that contain the largest connectivity degrees in the network, such sub sets of nodes can be chosen by a clustering algorithm or other preferred selection method. Comparing the mutual entropy of those selected clusters to the mutual entropies of random open networks with the same size shows that the clusters which possess most of the structural features of the network tend to have large differences in the entropies of different q-orders (with values that can be larger than those of the whole network), while clusters formed by small sub structures tend to show very small differences in the same quantity (lower than the average of the open networks), in agreement with the fact that the mutual entropies of clusters with few important structural properties should not vary much between different $q$-orders. Therefore, the use of mutual information in different q-orders is a promising tool to analyze the structure of a large network without requiring the knowledge of the totality of the information contained in it. The applications of this analysis in real world networks could provide a way to simplify or swift the current network structural analysis by eliminating the requirement of full knowledge of the network that is being analyzed; it can also be used to study network evolution by measuring the entropy changes.

## ACKNOWLEDGMENTS

This work was supported by DARPA through AFRL grant FA8750-04-2-0260.